\documentclass[11pt,twoside]{article}

\usepackage{mathtools} %Für := (\coloneqq)

\usepackage{mathrsfs} % used for mathscr

\newcommand{\FF}{\mathcal{F}}
\newcommand{\HH}{\mathscr H}

\usepackage{amsmath,amssymb,amsthm}
\usepackage{pstricks,pst-node,pst-coil,pst-plot,pstricks-add}
\usepackage{geometry,epsfig}
\usepackage{bbm}
\usepackage{cite}

\usepackage{cite}
\usepackage{paralist}
\usepackage{cleveref}
\usepackage{enumitem}
\usepackage{emptypage}

\DeclareGraphicsExtensions{.pdf}

\newtheorem{Lemma}{Lemma}[section]
\newtheorem{Theorem}[Lemma]{Theorem}
\newtheorem{Corollary}[Lemma]{Corollary}

\newcommand{\C}{\mathbb{C}} % komplexe
\newcommand{\R}{\mathbb{R}} % reelle
\DeclareMathOperator*{\esssup}{ess\,sup}

\DeclareRobustCommand{\rchi}{{\mathpalette\irchi\relax}}
\newcommand{\irchi}[2]{\raisebox{\depth}{$#1\chi$}} %richtiges Chi
\title{\textbf{Pointwise Bounds on Confined States\\ in Non-relativistic QED}}

\author{M.~Griesemer\footnote{marcel.griesemer@mathematik.uni-stuttgart.de}\,\ and V.~Ku{\ss}maul\footnote{valentin.kussmaul@mathematik.uni-stuttgart.de}\\  
\small Fachbereich Mathematik, Universit\"at Stuttgart, D-70569 Stuttgart, Germany}  
\date{}

\begin{document}
\maketitle

\begin{abstract}
Kato's well known distributional inequality for the magnetic Laplacian holds equally in the more general setting of non-relativistic quantum electrodynamics (QED), where the wave function is vector-valued and the vector potential is quantized. We give two new applications of this result: First, we show that eigenstates satisfy a subsolution estimate. Second, for general states, with energy distribution strictly below the ionization threshold, we give a short proof of pointwise exponential decay in the electronic configuration. 
\end{abstract}

\section{Introduction}

States of an atom or molecule with energy distribution strictly below the ionization threshold are well localized in a neighborhood of the nuclei: both in non-relativistic quantum mechanics and in the standard model of non-relativistic QED (the Pauli-Fierz model), the wave function decays exponentially as the distance $|x|$ of the electronic configuration $x=(x_1, \ldots,x_N)\in \R^{3N}$ from the nuclear positions grows. This decay implies an effective screening of the positive nuclear charges and it plays an important role in the mathematical analysis of many-particle quantum systems \cite{LT1986, AnaSig,AnaLew}. The purpose of this paper is to show how to derive pointwise (exponential) bounds from given $L^2$-bounds.

There is a well known standard argument to prove that exponential decay below the ionization threshold holds in the \emph{averaged} sense that
\begin{equation}\label{L2-decay}
 \int e^{2 \beta |x|}|\psi(x)|^2\, dx <\infty
\end{equation}
for the wave function $\psi:\R^{3N}\to \HH'$ with $|\psi(x)|$ the norm of $\psi(x)\in\HH'$, $\HH'$ being the tensor product of spin and Fock space \cite{MS10,BFS98,G,Panati2009}. The rate of decay, $\beta>0$, is explicit and depends on the difference between ionization threshold and least upper bound on the energy distribution of $\psi$. Statement \eqref{L2-decay} is called an $L^2$-exponential bound for $\psi$.

A very efficient way for turning  \eqref{L2-decay}  into a pointwise bound with the same rate is to use the mapping property 
 \begin{align}  
	e^{\beta | \cdot |}e^{- t H}e^{- \beta | \cdot |}: L^2(\R^{3 N}; \, \HH') \rightarrow L^\infty(\R^{3 N}; \, \HH')
	\label{MAPPING PROPERTY}
\end{align} 
of the semigroup $e^{- t H}$ generated by the Hamiltonian $H$ of the systems \cite{DHSV, M16,HiMa22}. The combination of \eqref{L2-decay} -- understood as an operator bound -- and \eqref{MAPPING PROPERTY} immediately implies the desired pointwise decay
\begin{equation}\label{L-infty-bound}
     \esssup_{x\in \R^{3 N}} e^{\beta |x|}|\psi(x)| <\infty.
\end{equation}
The trouble with this argument is the proof of \eqref{MAPPING PROPERTY}. It is typically done by means of a Feynman-Kac formula, which needs to be reestablished for every other model. This tends to be a lot of work, in particular for models from quantum field theory  \cite{HiMa22, M16, Hiro2019,HiHi2010, Hi14}.  
Below, after the following digression on subsolution estimates, we give a new, short argument for proving \eqref{MAPPING PROPERTY} in models from non-relativistic quantum field theory.

For eigenstates of Schr\"odinger operators there is the following alternative method for turning \eqref{L2-decay} into \eqref{L-infty-bound}
\cite{A}: Suppose $\psi$ is a solution to the magnetic Schr\"odinger equation
\begin{align*} 
	\left[ (-i \nabla + A)^2 + V \right] \psi = E \psi.
\end{align*} 
In view of Kato's distributional inequality \cite{Kato72},
\begin{align}\label{Kato}
   -\Delta | \psi | &\leq \mathrm{Re}\big( \overline{\mathrm{sgn} \, \psi} \, (-i \nabla + A)^2 \psi\big),
\end{align}
where $\mathrm{sgn} \, \psi = \psi / |\psi|$ ($\mathrm{sgn} \, \psi(x) = 0$ if $\psi(x) = 0$), it follows that $|\psi|$ is a subsolution to a non-magnetic Schr\"odinger equation of the form
\begin{equation} \label{def:subsolution}
  (-\Delta + V) | \psi | \leq E |\psi|.
\end{equation} 
Under weak assumptions on the negative part of $V$, \eqref{def:subsolution} implies a subsolution estimate
\begin{equation}\label{Agmon}
    \esssup_{y\in B(x,1/2)}|\psi(y)| \leq C\bigg(\int_{B(x,1)}|\psi(y)|^2\, dy\bigg)^{1/2}
\end{equation}
for all $x\in \R^{3 N}$ \cite{A,AiSi, Simader90}. It follows that $L^2$-exponential decay, such as \eqref{L2-decay}, with a Lipschitz function $x \mapsto \beta|x|$ implies the corresponding pointwise decay \eqref{L-infty-bound}. This line of arguments is easily generalized to eigenstates of the Pauli-Fierz hamiltonian, $H$, of non-relativistic QED: from a generalization of Kato's distributional inequality \eqref{Kato} to vector-valued wave functions and quantized vector potentials we see that, in a distributional sense,
\begin{align}  
	S |\psi| \leq \mathrm{Re} \, (\mathrm{sgn} \, \psi, H \psi),
	\label{intro: Kato type S H ineq}
\end{align} 
for any $\psi$ belonging to the domain of $H$. Here $S = -\Delta + V + \mathrm{const}$ with a constant resulting from the coupling of the electronic spin to the quantized magnetic field. As in \eqref{L2-decay}, $|\psi(x)|$ denotes the norm of $\psi(x) \in \HH'$, and $(\cdot, \cdot)$ is the inner product in $\HH'$. If $H \psi = \lambda \psi$, then \eqref{intro: Kato type S H ineq} implies $S|\psi| \leq \lambda |\psi|$, which is an inequality of the form \eqref{def:subsolution}. As a direct consequence, we obtain the subsolution estimate \eqref{Agmon} for the eigenstate $\psi$. In particular, any $L^2$-exponential bound implies the corresponding pointwise bound. 

We now return to the problem of proving \eqref{MAPPING PROPERTY}. By a generalization of Simon's comparison theorem \cite{Si79} -- see \Cref{Simon's theorem}, below -- the Kato type inequality \eqref{intro: Kato type S H ineq} implies for all $\psi \in L^2(\R^{3 N}; \, \HH')$ the pointwise inequality 
\begin{equation}\label{semi-group-bound}  
|e^{-t H}\psi | \leq e^{-t S} |\psi|.
\end{equation}
It therefore suffices to prove the analogue of \eqref{MAPPING PROPERTY} for the Schr\"odinger semigroup $e^{-t S}$. This, in fact, is a well known result established first by a Dyson-Phillips expansion \cite{DHSV} and little later by the well known Feynman-Kac representation for Schr\"odinger semigroups \cite{Si82}. So by \eqref{semi-group-bound}, the problem of proving \eqref{MAPPING PROPERTY} is reduced to a solved one and there is no need of a Feynman-Kac representation for a quantum field theory semigroup. A fortiori analogous results hold for the UV-regularized Nelson model. Due to domain issues our results cannot be readily applied to the renormalized Nelson model. But we expect that, with some more work, our methods should also allow one to establish pointwise bounds for that model, see \cite{HiMa22}.

The mapping property \eqref{MAPPING PROPERTY} in non-relativistic QED is due to Matte \cite{M16}. His result, as all other previous results concerning pointwise bounds in models of quantum field theory (QFT), to our knowledge, are based on Feynman-Kac formulas  \cite{HiMa22, M16, Hiro2019,HiHi2010, Hi14}. Prior to \cite{M16}, Hidaka and Hiroshima \cite{HiHi2010} proved pointwise exponential decay of eigenvectors by adapting Carmona's estimates on Feynman-Kac integrals for Schr\"odinger operators. Their method does not depend on the $L^2$-decay \eqref{L2-decay}, but leads to suboptimal decay rates. 
The advantage of the Feynman-Kac formula established in \cite{M16} is that it reveals further mapping properties of the semigroup $e^{-t H}$, finer 
than \eqref{MAPPING PROPERTY}, concerning  e.g. the decay of $\psi(x)\in\HH'$, for fixed $x$, in the photonic degrees of freedom, see Theorem 5.2 and  
Example 8.3 in \cite{M16}. Moreover, it is shown in \cite[Theorem 8.1]{M16} that $x \mapsto \psi(x) \in \HH'$ is continuous, which means that  ``$\esssup$'' in \eqref{L-infty-bound} may be replaced by ``$\sup$''. We note that in the $N = 1$ electron case this also follows from the fact that $H^2(\R^3; \, \HH')$-functions are (H\"older) continuous due to a Sobolev embedding theorem. H\"older continuity alone can be used to derive \emph{some} pointwise exponential decay from \eqref{L2-decay}, but the rate so obtained is worse than in \eqref{L2-decay}, see \cite{Ahlrichs}. For results on $L^2$-exponential decay in other models of QFT we refer to \cite{MS10, Panati2009}. For the existence of eigenstates in the Pauli-Fierz model see \cite{BFS,GLL,LL} and the references therein. 

The generalized version of Kato's distributional inequality that we present in \Cref{Kato sec}, in the case of QED, was previously established in \cite[Theorem 3.2]{KMS13}. We provide a simple and independent proof. Kato's distributional inequality was originally conceived to prove essential self-adjointness for (magnetic) Schr\"odinger operators with singular potentials \cite{Kato72}. For the inequality to be useful to that end, it must be established under minimal assumptions \cite[Lemma A]{Kato72}. In the context of QED, this seems difficult. By Simon's comparison theorem \cite{Si79}, inequalities of Kato type are equivalent to pointwise inequalities between semigroups. In the present paper, see \Cref{Simon's theorem}, Simon's comparison theorem is generalized to a vector-valued setting.

This paper is organized as follows. \Cref{Kato sec} is written in an abstract setting and contains the generalizations of Kato's distributional inequality, \Cref{Kato ineq},  and Simon's comparison theorem, \Cref{Simon's theorem}. The Pauli-Fierz model is introduced in \Cref{sec:qft}. In \Cref{sec:qft-exp} we establish subsolution estimates for eigenstates, see \Cref{thm:ptw-bound}. Pointwise exponential decay for confined states is proved in \Cref{sec:confined}.

%---------------------------------------------------------------------------------------------------------------------------------------------------------------------

\section{Kato's distributional inequality}
\label{Kato sec}
This section collects the abstract results on which our method is based. The first theorem is a generalization of Kato's distributional inequality \cite{Kato72} to quantized vector potentials. The second theorem is a generalization of a theorem about the equivalence of a Kato type inequality and a pointwise inequality between semigroups \cite{Si79}.

Let $\HH'$ be a separable Hilbert space with inner product $(\cdot, \cdot)$ and norm $| \cdot |$. Let $\HH$ be the Hilbert space $L^2(\R^n; \, \HH')$. The inner product in $\HH$ is denoted with 
\begin{align*}  
\langle \varphi, \psi \rangle &= \displaystyle \int_{\R^n} (\varphi(x), \psi(x)) \, dx, \quad \varphi, \psi \in \HH 
\end{align*}  
and the norm with $\| \cdot \|$. For each $m = 1, ..., n$ let the operator $A_m$ be given as a direct integral of (possibly unbounded) symmetric operators $(A_m(x))_{x \in \R^n}$ acting on $\HH'$, that is $A_m$ acts on $\HH$ according to
\begin{align*} 
(A_m \psi)(x) = A_m(x) \psi(x), \quad x \in \R^n. 
\end{align*} 
We denote by $p_m = - i \partial_{m}$ the $m$-th component of the momentum operator, defined on it's natural domain of self-adjointness. Moreover, we set
\begin{align*}
\psi \in D(p + A) &\overset{Def.}{\iff} \psi \in D(p_m) \cap D(A_m)  \: \,  \forall m = 1, ..., n.\\
\psi \in D((p + A)^2) &\overset{Def.}{\iff} \psi \in D(p + A) \, \, \text{and} \, \, p_m \psi + A_m \psi \in D(p_m) \cap D(A_m) \: \,  \forall  m = 1, ..., n. 
\end{align*}

\begin{Theorem}
\label{Kato ineq}
For $\psi \in \HH$ define 
\begin{align*}
(\mathrm{sgn} \, \psi)(x) =
\begin{dcases}
\dfrac{\psi(x)}{|\psi(x)|} \quad & \psi(x) \neq 0, \\
\, \, \, 0 \quad & \psi(x) = 0.
\end{dcases}
\end{align*}

\begin{enumerate}
\item[(i)] \textbf{Diamagnetic inequality}: If $\psi \in D(p + A)$, then $| \psi |$ is weakly differentiable and 
$$\nabla | \psi | = \mathrm{Re} \, ( \mathrm{sgn} \,  \psi, (\nabla + i A) \psi). $$
In particular $| \nabla | \psi | | \leq | (\nabla + i A) \psi |$ and $| \psi | \in H^1(\R^n)$. 
\item[(ii)] \textbf{Kato's distributional inequality}:  If $\psi \in D((p + A)^2)$, then
\begin{align*} 
-\Delta | \psi | \leq \mathrm{Re} \, (\mathrm{sgn} \, \psi, (-i \nabla + A)^2 \psi)
\end{align*} 
in the sense of distributions on $C_0^\infty(\R^n)$. 
\end{enumerate}
\end{Theorem}

\begin{proof} 
The following proof is inspired by Kato's original proof \cite[Lemma A]{Kato72}, which concerns the case $\HH'=\C$. 

(i) From $\psi \in D(p + A) \subset D(p)$, it follows that $|\psi|^2 = (\psi, \psi)$ is weakly differentiable and 
\begin{align}  
\nabla | \psi |^2 = (\nabla \psi, \psi) + (\psi, \nabla \psi) = 2 \mathrm{Re} \, (\psi, \nabla \psi).
\label{e0}
\end{align} 
For $\varepsilon > 0$ we define $\psi_\varepsilon = ( | \psi |^2 + \varepsilon^2)^{1/2}$.  By the chain rule (Lemma \ref{Kettenregel}), we see that
\begin{align}
\nabla \psi_\varepsilon &= \dfrac{1}{2} ( | \psi |^2 + \varepsilon^2)^{-1/2} \nabla | \psi |^2 = \mathrm{Re} \left( \dfrac{\psi}{\psi_\varepsilon}, \nabla \psi \right).
\label{e1}
\end{align}
Since $\psi / \psi_\varepsilon \rightarrow \mathrm{sgn} \, \psi$ pointwise and $| \psi / \psi_\varepsilon | \leq 1$, it follows that $\nabla \psi_\varepsilon \rightarrow \mathrm{Re} (\mathrm{sgn} \, \psi, \nabla \psi)$ in $L^2(\R^n)$. In conjunction with $\psi_\varepsilon \rightarrow | \psi |$ in $L_\mathrm{loc}^2(\R^n)$, it follows that $| \psi |$ is weakly differentiable and 
$$\nabla | \psi | = \mathrm{Re} (\mathrm{sgn} \, \psi, \nabla \psi) = \mathrm{Re} (\mathrm{sgn} \, \psi, (\nabla + i A) \psi),$$ 
where the symmetry of the operators $A_{m}(x)$ was used in the last equation.

(ii) From part (i) it follows that $|\psi| \in H^1(\R^n)$ and hence $\nabla |\psi|^2 = 2 |\psi|  \nabla | \psi |$. Therefore Equation \eqref{e1} combined with \eqref{e0} now becomes 
\begin{align*}  
\nabla \psi_\varepsilon = \frac{| \psi | \nabla | \psi|}{\psi_\varepsilon},
\end{align*} 
and, by the chain rule (\Cref{Kettenregel}), we arrive at
\begin{align}  
\nabla \frac{1}{\psi_\varepsilon} = - \frac{| \psi | \nabla | \psi |}{\psi_\varepsilon^3}.
\label{e3}
\end{align} 
The assumption $\psi \in D((p + A)^2)$ implies that $(\psi, (\nabla + i A)\psi)$ is weakly differentiable, and, by the symmetry of $A(x)$, that 
\begin{align}  
\nabla \cdot (\psi, (\nabla + i A) \psi) &= (\nabla \psi, ( \nabla + i A) \psi ) + ( \psi, \nabla \cdot (\nabla + i A) \psi)\nonumber \\ 
&= ( (\nabla + i A) \psi , (\nabla + i A) \psi)  + ( \psi, (\nabla + i A) \cdot (\nabla + i A) \psi)\nonumber \\
&= | (\nabla + i A) \psi |^2 + ( \psi, (\nabla + i A)^2 \psi).\label{p-and-A}
\end{align} 
After these preparations we are now ready to prove part (ii): from part (i) - or \eqref{e0} and the symmetry of $A(x)$ - we know that
\begin{align*}  
| \psi | \nabla |\psi| = \mathrm{Re} \, (\psi, (\nabla + i A) \psi).
\end{align*} 
Differentiating this with the help of \eqref{p-and-A} we arrive at the key identity
\begin{align}  
\nabla \cdot ( | \psi | \nabla |\psi| ) =  | (\nabla + i A) \psi |^2 + \mathrm{Re} \, ( \psi, (\nabla + i A)^2 \psi).
\label{key}
\end{align} 
From  \eqref{e3}, \eqref{key} and the product rule (\Cref{Produktregel}), it follows that
\begin{align}  
\nabla \cdot \left( \frac{1}{\psi_\varepsilon} | \psi | \nabla | \psi| \right) &= \nabla \left( \frac{1}{\psi_\varepsilon} \right) \cdot | \psi|\nabla | \psi | +  \frac{1}{\psi_\varepsilon} \nabla \cdot ( | \psi| \nabla | \psi|) \nonumber \\
&= - \frac{| \psi |^2 |\nabla | \psi ||^2}{\psi_\varepsilon^3} +  \frac{1}{\psi_\varepsilon}  \Bigl( | (\nabla + i A) \psi |^2 + \mathrm{Re} \, ( \psi, (\nabla + i A)^2 \psi) \Bigr) \nonumber \\
&\geq \frac{1}{\psi_\varepsilon} \mathrm{Re} \, ( \psi, (\nabla + i A)^2 \psi).
\label{e4}
\end{align} 
In the last line we used $| \psi|^2 / \psi_\varepsilon^2 \leq 1$ and $|\nabla |\psi||^2 \leq | (\nabla + i A) \psi|^2$. We now integrate \eqref{e4} against a non-negative function $\phi \in C_0^\infty(\R^n)$ and obtain
\begin{align*}  
- \int   \nabla \phi  \cdot \frac{|\psi|}{\psi_\varepsilon} \nabla |\psi| \, dx \geq \int \phi \, \mathrm{Re}  \left( \frac{\psi}{\psi_\varepsilon}, (\nabla + i A)^2 \psi \right) dx.
\end{align*} 
In the limit $\varepsilon \rightarrow 0$ we obtain, by dominated convergence, 
\begin{align*}  
- \int\nabla \phi \cdot \nabla |\psi| \, dx \geq \int    \phi \, \mathrm{Re}  \left( \mathrm{sgn} \, \psi, (\nabla + i A)^2 \psi \right)  dx.
\end{align*} 
 This concludes the proof of (ii). 
\end{proof}

The following theorem, in the case $\HH' = \C$, is due to Simon \cite{Si79}. The proof is short and easily generalized to the present vector-valued setting.

\begin{Theorem} 
Let $H = H^{*}$ be semibounded acting on $\HH = L^2(\R^n; \, \HH')$.
Let $S = S^{*}$ \hspace{-3mm} be semibounded acting on $L^2(\R^n)$ and suppose that $S$ is the generator of a positivity preserving semigroup. Then the following statements are equivalent:
\begin{enumerate}
\item[(i)] \textbf{Kato inequality:} If $\psi \in D(H)$, then $|\psi| \in Q(S)$ and for all $0 \leq \phi \in Q(S)$ we have
\begin{align}  
\langle \phi, S |\psi| \rangle \leq \mathrm{Re} \, \langle \phi \,  \mathrm{sgn} \, \psi, H \psi \rangle.
\label{Kato-type inequality}
\end{align} 
Here $\langle \cdot, S \cdot \rangle$ is understood in form sense. 
\item[(ii)] \textbf{Semigroup inequality:} For all $\psi \in L^2(\R^n; \, \HH')$ and $t > 0$ we have 
\begin{align*}  
	| (e^{-t H} \psi)(x) | \leq (e^{-t S} |\psi|)(x) \quad \text{for a.e. } x \in \R^n. 
\end{align*} 
\end{enumerate}
\label{Simon's theorem}
\end{Theorem}

To verify the hypotheses about $S$ in the theorem it is useful to recall from \cite{Si77}, that the semigroup generated by a semibounded self-adjoint operator $S$ is positivity preserving if and only if, 
first, the Kato inequality (i) holds with $H=S$, and second, $\psi \in Q(S)$ implies $|\psi| \in Q(S)$. Since these conditions are satisfied for $S = -\Delta$, they also hold for $S = -\Delta + V$ with $V$ real-valued and infinitesimally operator- (hence form-)  bounded with respect to $-\Delta$. For such potentials, we thus conclude that  $S = -\Delta + V$ is the generator of a positivity preserving semigroup. The operator $H$ in our application will be the Hamiltonian defined in the next section.

%%%%%%%%%%%%%%%%%%%%%%%%%%%%%%%%%%%%%PAULI-FIERZ%%%%%%%%%%%%%%%%%%%%%%%%%%%%%%%%%%%%%%%%%%%%%%%%%%%%%%%%%%%%

\section{The Pauli-Fierz model}
\label{sec:qft}

We now introduce the Pauli-Fierz model for $N$ identical spin-$1/2$ particles, called electrons, in $\R^3$. 
More elaborate descriptions may be found in \cite{BFS,Gri2006,HH}.

The one-particle Hilbert space associated with the photon field is $\mathfrak{h} = L^2(\R^3 \times \{1, 2\})$. The symmetric Fock space over $\mathfrak{h}$ is denoted by
\begin{align*} 
    \mathcal{F}^{+} =\bigoplus_{n = 0}^\infty \otimes_\mathrm{sym}^n \,  \mathfrak{h},\qquad \otimes_\mathrm{sym}^0 \,  \mathfrak{h} := \C.
\end{align*}
For $h \in \mathfrak{h}$ let $a(h)$ and $a^*(h)$ denote the usual bosonic annihilation and creation operators in $\mathcal{F}^{+}$, and let
\begin{align*} 
\phi(h) = a(h) + a^*(h),
\end{align*} 
which is defined and symmetric on the subspace of finite particle vectors from $\mathcal{F}^{+}$. The closure of this operator is self-adjoint and denoted by the same symbol.
Let
\begin{align*} 
       \omega(k) = | k |, \quad k \in \R^3,
\end{align*} 
be the photon dispersion relation and let $H_f = d\Gamma(\omega)$ be the second quantization of multiplication with $\omega$ in $\mathfrak{h}$. 

The Hilbert space of the model (without statistics yet) is the tensor product
\begin{align}
     \HH = L^2(\R^{3 N}) \otimes \bigg(\bigotimes_{j = 1}^N \C^2\bigg) \otimes \mathcal{F}^{+}
\label{full Hilbert space}
\end{align}
of particle and Fock space $\FF^{+}$, with the $N$ factors of $\C^2$ accounting for the spin degrees of the particles. The inner product and norm of $\HH$ will be denoted by $\langle \cdot, \cdot \rangle$ and $\| \cdot \|$, respectively. 

The Pauli-Fierz Hamiltonian is composed of operators acting on the various factors of  \eqref{full Hilbert space}. The operator $H_f = d\Gamma(\omega)$ in $\FF^{+}$ accounts for the energy of massless photons. The potential $V:\R^{3 N} \rightarrow \R$ acts by multiplication on the first factor of \eqref{full Hilbert space}. We
assume that
\begin{align}
V(x) &= \sum_{j = 1}^N v(x_j) + \sum_{j < k} w(x_j - x_k), \quad x = (x_1, ..., x_N) \in \R^{3 N}, \label{potential}
\\
&v,w \in L^2(\R^3) + L^{\infty}(\R^3), \qquad w(x) = w(-x). \nonumber
\end{align}
It follows that $V$ is infinitesimally operator- (and hence form-) bounded with respect to the Laplacian $-\Delta$. As usual we shall not distinguish in notation between the operator $V$ in $L^2(\R^{3 N})$ and the operator $V \otimes 1$ in $\HH$. The same remark applies to $-\Delta$ and the components of the momentum operator $- i \nabla_j$ of the $j$th electron.  

To define quantized vector potential and magnetic field we introduce, for $x  \in \R^3$ and $\ell = 1,2,3$, the elements
$G_{\ell}(x), F_{\ell}(x) \in \mathfrak{h}$ by the functions
\begin{align}
  [G_{\ell}(x)](k, \lambda) &= \sqrt{\alpha} \frac{\rchi_\Lambda(k)}{\sqrt{\omega(k)}} e^{-i k \cdot x} \varepsilon_\ell(k, \lambda),  \label{def G}\\
[F_{\ell}(x)](k, \lambda) &= -i  \sqrt{\alpha} \frac{\rchi_\Lambda(k)}{\sqrt{\omega(k)}} e^{-i k \cdot x} (k \wedge \varepsilon(k, \lambda))_\ell, \quad k \in \R^3, \lambda = 1,2. \label{def F}
\end{align}
Here $\Lambda < \infty$ is an arbitrary ultraviolet cutoff and $\rchi_\Lambda$ denotes the characteristic function of the set $\{|k| \leq \Lambda \}$. The coupling constant $\alpha > 0$ is also arbitrary. The polarization vectors  $\varepsilon(k, 1), \varepsilon(k, 2)\in \R^3$, for $k\neq 0$, are normalized and orthogonal to $k \in \R^3$. In fact, our results do not depend on the exact structure of \eqref{def G} and \eqref{def F}. For example, the cutoff function $\rchi_\Lambda$ could also be chosen from Schwartz space $S(\R^3)$ and expression \eqref{def F} could contain a factor of $g/2$ with $g \in \R$ to account for the anomalous gyromagnetic ratio of the electron. 

With the identification $L^2(\R^{3 N}) \otimes \FF^+ \, \widetilde{=} \, L^2(\R^{3 N} ; \FF^+)$ we can define
\begin{align}
(A_{j, \ell} \, \psi)(x) &= \phi(G_\ell(x_j)) \psi(x), \nonumber \\ 
(B_{j, \ell} \, \psi)(x) &= \phi(F_\ell(x_j)) \psi(x).
\label{vector potential}
\end{align}
By $A_j$ and $B_j$ we denote the operator-valued vectors with components
$A_{j, \ell}$ and $B_{j, \ell}$, $\ell = 1,2,3$. With the above notations and conventions the Hamiltonian of the system reads
\begin{equation}
   H=\sum_{j = 1}^N \left[ (-i \nabla_j +   A_j)^2 + \, \sigma_j \cdot B_j \right]  + V + H_f,
\label{HAMILTONIAN}
\end{equation}
where $\sigma_j$ denotes the triple of Pauli matrices acting on the $j$th factor in $\bigotimes_{j = 1}^N \C^2$. 
It is well known that for all values of the coupling constant $\alpha > 0$ the Hamiltonian is self-adjoint on $D(H) = D(-\Delta + H_f) = D(-\Delta) \cap D(H_f)$ and bounded from below \cite{Hi2002,HH}. Moreover, $H$ is essentially self-adjoint on any core for $-\Delta + H_f$ \cite{HH}.

\medskip
To fit the Pauli-Fierz model into the setting of the previous section, we now define
\begin{align*}  
    \HH'= \bigg(\bigotimes_{j = 1}^N \C^2 \bigg) \otimes \mathcal{F}^{+}
\end{align*} 
with inner product $( \cdot , \cdot )$ and norm $| \cdot |$. Then $\HH$ can be identified with $L^2(\R^{3 N};\, \HH')$ via the unitary map that is determined by $\psi = f \otimes v \mapsto \psi(x) = f(x) \, v$. 
To further simplify notation we introduce operator-valued vectors $A,B$ and $\sigma$ with $3N$ components
such that \eqref{HAMILTONIAN} becomes
\begin{equation}\label{Pauli-Fierz} 
     H = (-i \nabla +  A)^2 +  \sigma \cdot B + V + H_f.
\end{equation} 

To account for the fermionic nature of electrons, we now reduce the Hilbert space to the closed subspace $\HH^{-}\subset \HH$ consisting of all vectors $\psi$ that are antisymmetric with respect to permutations of the $N$ factors of $L^2(\R^3)\otimes \C^2$ in \eqref{full Hilbert space}. By the above mentioned isomorphism, $\HH^{-}$ may be identified with a closed subspace of $L^2(\R^{3 N};\, \HH')$.  
Since $H$ is symmetric w.r.t.~permutations of the $N$ electrons, it follows that the orthogonal projection onto $\HH^{-}$ commutes with $H$. We thus can restrict $H$ onto $\HH^{-}$ and obtain a self-adjoint operator
\begin{align*}
H^{-} = H: D(H) \cap \HH^{-} \rightarrow \HH^{-}
\end{align*} 
called the \emph{Pauli-Fierz Hamiltonian}. The \emph{ionization threshold} of $H^-$ is defined by 
\begin{equation}
\label{def:sigma}
\Sigma = \lim_{R \rightarrow \infty} \left( \inf_{\psi \in D_R} \langle \psi, H^- \psi \rangle \right)
\end{equation}
where $D_R\subset D(H^-)$ is the subset of normalized states supported outside the ball $B(0,R) \subset \R^{3N}$ \cite{G}. 

%The \textbf{ionization energy} is the energy difference $\Sigma - \inf \sigma(H^{-})$. In the case of many-particle potentials of the form \eqref{potential}, it is well known that, see Theorem 3 in \cite{G}, 
%\begin{align*}
 %\Sigma = \min_{N' = \, 1, ..., N} \left( E_{N - N'} + E_{N'}^0 \right).
%\end{align*}
%Here $E_N = \inf \sigma(H^-)$ is the lowest energy of the $N$-electron system and $E_{N  = \, 0} = 0$. $E_N^0$ is the lowest energy of the $N$-electron system, if the external potentials are dropped in (\ref{potential}).  

States with energy distribution strictly below $\Sigma$ decay exponentially in the following sense \cite[Theorem 1]{G}:
For $\lambda \in \R$ let
 \begin{align*}  
 P_\lambda^{-} \coloneqq \rchi_{(- \infty, \lambda]}(H^{-}).
 \end{align*} 
 If $\lambda < \Sigma$ and $\beta < \sqrt{\Sigma - \lambda}$, then
 \begin{align}  
 \| e^{\beta | \cdot |}P_\lambda^{-} \| < \infty.
 \label{L 2 operator localization}
 \end{align} 
 In particular, for all $\psi \in \mathrm{Ran} \, P_\lambda^{-}$
\begin{align}
        \int  e^{2 \beta |x|} | \psi(x) |^2 dx < \infty.
\label{L 2 exponential decay}
\end{align}
This includes eigenvectors of $H^{-}$ with eigenvalue $\lambda$ below $\Sigma$ in which case the result is from \cite[Lemma 6.2]{GLL} and the proof is easier.

%------------------------------------------------------------------------------------------------------------------------------SUBSOLUTION ESTIMATES FOR EIGENSTATES---------------------------------------------

\section{Subsolution estimates for eigenstates}
\label{sec:qft-exp}

This section is devoted to \emph{local} pointwise bounds on eigenstates $\psi$ of the Pauli-Fierz Hamiltonian. That is, local $L^\infty$-norms of $|\psi|$ are estimated in terms of local $L^2$-norms.
When combined with the known $L^2$-exponential bound \eqref{L 2 exponential decay}, a pointwise exponential bound with the same rate of decay follows. Pointwise exponential decay for more general  states is the subject of the next section.

First we show that a Kato-type inequality holds between a Schr\"odinger operator and the Hamiltonian \eqref{Pauli-Fierz}, see \Cref{S H Kato ineq} below. It then follow that $|\psi|$ is a subsolution to a Schrödinger equation, which, by a standard result leads to a subsolution estimate for $|\psi|$ \cite{A}. 

\begin{Theorem}
\label{S H Kato ineq}
Let $H$ be the Hamiltonian \eqref{Pauli-Fierz}. We define the Schrödinger operator
\begin{align}  
S \coloneqq -\Delta + V - \frac{8 \pi}{3} \alpha N^2 \Lambda^3.
\label{definition S}
\end{align} 
If $\psi \in D(H)$, then $| \psi | \in H^1(\R^{3 N})$ and
\begin{align}  
	S |\psi| \leq \mathrm{Re} \, (\mathrm{sgn} \, \psi, H \psi)
	\label{S H dist ineq}
\end{align} 
in the sense of distributions on $C_0^\infty(\R^{3 N})$. 
\end{Theorem}
\noindent \emph{Remark:} If electron spin is neglected, then \eqref{S H dist ineq} holds with $S = -\Delta + V$. 
\begin{proof}
Let $\psi \in D(H)$. Since $D(H) \subset D((p + A)^2)$, see \cite[Theorem 7]{HH}, \Cref{Kato ineq} implies that $|\psi| \in H^1(\R^{3 N})$ and
\begin{align*}  
 -\Delta | \psi | \leq \mathrm{Re} \, (\mathrm{sgn} \, \psi, (-i \nabla + A)^2 \psi)
\end{align*} 
in the sense of distributions on $C_0^\infty(\R^{3 N})$. Moreover, we have pointwise on $\R^{3 N}$
\begin{align}  
V |\psi| &= \mathrm{Re} \, (\mathrm{sgn} \, \psi, V \psi), \nonumber \\
- \frac{8 \pi}{3} \alpha N^2 \Lambda^3 |\psi| &\leq \mathrm{Re} \, (\mathrm{sgn} \, \psi, (H_f + \sigma \cdot B) \psi).
\label{ineq: spin lower bound}
\end{align} 
Inequality \eqref{ineq: spin lower bound} follows from \Cref{lower bound on spin}. Adding the above inequalities concludes the proof. If electron spin is neglected, then \eqref{ineq: spin lower bound} can be replaced with $0 \leq \mathrm{Re} \, (\mathrm{sgn} \, \psi, H_f \, \psi)$. 
\end{proof}

\begin{Theorem}
\label{thm:ptw-bound}
If $\psi$ is an eigenvector of the Hamiltonian \eqref{Pauli-Fierz}, then $| \psi |$ satisfies a subsolution estimate: For any positive real numbers $r,R \in \R$, with $r < R$, there exists a constant $C$ such that for all $x \in \R^{3 N}$
\begin{align}
  \esssup_{y\in B(x,r)}|\psi(y)| \leq C\bigg(\int_{B(x,R)}|\psi(y)|^2\, dy\bigg)^{1/2}.
\label{main bound}
\end{align}
\end{Theorem}

\begin{proof}
If $H \psi = \lambda \psi$, then \Cref{S H Kato ineq} implies $|\psi| \in H^1(\R^{3 N})$ and 
\begin{align*}  
S |\psi| \leq \mathrm{Re} \, (\mathrm{sgn} \, \psi, H \psi) = \lambda | \psi |.
\end{align*} 
Denoting with $V_{-} = \mathrm{max} (0, -V)$ the negative part of the potential $V$, it follows that $|\psi|$ is a subsolution to a Schrödinger equation of the form
\begin{align}  
- \Delta |\psi | \leq \left(V_{-} + \frac{8 \pi}{3} \alpha N^2 \Lambda^3 + \lambda  \right) |\psi|.
\label{subsolution ineq}
\end{align} 
By Theorem 5.1 in Agmon's book \cite{A}, the subsolution estimate \eqref{main bound} follows. Actually, Theorem 5.1 in Agmon's book is about solutions rather than subsolutions to  Schr\"odinger equations, but the first step in the proof is to derive an inequality of the form \eqref{subsolution ineq} and the rest follows from this inequality. The class of potentials considered by Agmon is more general than  \eqref{potential}.
\end{proof}

Subsolution estimates of the form  \eqref{main bound} for ($\C$-valued) subsolutions to Schr\"odinger equations were known previous to Agmon's work, see \cite{A, AiSi, Simader90} and the references therein, but for the 
proof of \Cref{thm:ptw-bound} the version of Agmon appears most convenient. %In \cite[Theorem 6.1]{AiSi} local boundedness is assumed of a subsolution. 

%---------------------------------------------------------------------------------------------------------------------------------

%---------------------------------------------------------------------------------------------------------------------------------
Theorem \ref{thm:ptw-bound} allows us to pass from $L^2$ exponential bounds to $L^\infty$ exponential bounds:

\begin{Corollary}
\label{anisotrope Schranken}
Let $\psi$ be an eigenvector of the Hamiltonian~\eqref{Pauli-Fierz} and suppose that
\begin{align*} 
\int  e^{2 f(x)}  |\psi(x)|^2 dx< \infty, 
\end{align*} 
with some Lipschitz function $f : \R^{3 N} \rightarrow [0, \infty)$. Then 
\begin{align*} 
\esssup_{x \in \R^{3 N}}  \, e^{f(x)}  |\psi(x)| < \infty.
\end{align*} 
\end{Corollary}

\begin{proof} 
Let  $r = 1/2$, $R = 1$ and let $C$  be given by Theorem \ref{thm:ptw-bound}. Denoting by $L$ the Lipschitz constant of $f$, it follows that
\begin{align*} 
\esssup_{y \, \in B(x, 1/2)} e^{f(y)} | \psi(y) | &\leq C e^{L/2} e^{f(x)} \bigg(\int_{B(x,1)}|\psi(y)|^2\, dy\bigg)^{1/2} \\ 
&= C e^{L/2}  \bigg(\int_{B(x,1)}e^{2 f(x)} |\psi(y)|^2\, dy\bigg)^{1/2} \\ 
&\leq C e^{L/2} e^L \bigg(\int e^{2 f(y)} |\psi(y)|^2\, dy\bigg)^{1/2}
\end{align*} 
for all $x \in \R^{3 N}$. Note that the right-hand side does not depend on $x$.
\end{proof}
%---------------------------------------------------------------------------------------------------------------------------------------
\Cref{anisotrope Schranken} combined with the $L^2$-exponential bound \eqref{L 2 exponential decay} yields pointwise exponential decay for eigenstates of the Pauli-Fierz Hamiltonian. In the following section we will prove this result for more general states.

%------------------------------------------------------------------------------------------------------------------------------POINTWISE BOUNDS ON CONFINED STATES---------------------------------------------

\section{Pointwise bounds on confined states}
\label{sec:confined}

In this section we establish pointwise exponential decay for states of the Pauli-Fierz model with energy distribution strictly below the ionization threshold. This result is based on the $L^2$-exponential bound \eqref{L 2 operator localization} and the mapping properties of the semigroup of the Hamiltonian \eqref{Pauli-Fierz}, which we derive using the Kato inequality of \Cref{S H Kato ineq} and well-known mapping properties of Schrödinger semigroups \cite{DHSV, Si82}.

%----------------------------------------------------------------------------------------------------------------
\begin{Theorem}
\label{thm: QED semigroup inequality}
Let $H$ be the Hamiltonian \eqref{Pauli-Fierz} and $S$ the Schrödinger operator \eqref{definition S}.
Then for all $t > 0$ and $\psi \in \HH = L^2(\R^{3 N}; \, \HH')$ we have pointwise a.e. on $\R^{3 N}$
\begin{align}  
|e^{-t H}\psi | \leq e^{-t S} |\psi|.
\label{QED semigroup inequality}
\end{align} 
\end{Theorem}
%----------------------------------------------------------------------------------------------------------------
\noindent
\emph{Remark:} Given the Feynman-Kac representations for $e^{-t H}$ and $e^{-t S}$, \eqref{QED semigroup inequality} easily follows from the triangle inequality for (stochastic) integrals, see \cite{M16, H97}. Our proof below uses \Cref{Simon's theorem}. A third way of establishing \eqref{QED semigroup inequality}, at least for quantized vector potentials regularized in the infrared, is based on suitable gauge transformations and the Trotter product formula \cite{Hiro1996}.

\begin{proof}
The assertion agrees with statement (ii) of \Cref{Simon's theorem}. It remains to verify the hypotheses of that theorem for the present choices of $S$ and $H$. This is done as follows: By the discussion at the end of \Cref{Kato sec} the form domain of $S$ is $Q(S) = H^1(\R^{3 N})$ and the semigroup of $S$ is positivity preserving. From \Cref{S H Kato ineq} we know that $\psi \in D(H)$ implies $|\psi| \in H^1(\R^{3 N})$ and that 
\begin{align*}  
\langle \phi, S |\psi| \rangle \leq \mathrm{Re} \, \langle \phi \, \mathrm{sgn} \, \psi, H \psi \rangle,
\end{align*} 
for functions $0 \leq \phi \in C_0^\infty(\R^{3 N})$. Here $\langle \cdot, S \cdot \rangle$ is understood in form sense. By a simple approximation argument, this inequality extends to $0 \leq \phi \in H^1(\R^{3 N})$. We conclude that statement (i) of \Cref{Simon's theorem} holds and therefore, so does statement (ii). This concludes the proof.
\end{proof}

The meaning of \eqref{QED semigroup inequality} is that mapping properties of the Schr\"odinger semigroup are inherited by the semigroup of $H$. 
The mapping properties of the Schr\"odinger semigroup are those of the heat semigroup that extend to $e^{-tS}$ by means of a Dyson-Phillips expansion or the Feyman-Kac representation \cite{DHSV, Si82}. We recall from these papers that, for any non-negative Lipschitz function $f: \R^{3 N} \rightarrow \R$, the linear operator $e^{f}e^{-t S}e^{-f}$ is bounded from 
$L^2(\R^{3 N})$ to $L^{\infty}(\R^{3 N})$  -- in fact, from $L^p(\R^{3 N})$ to $L^{q}(\R^{3 N})$ with  $1 \leq p < q \leq \infty$.  The $\HH'$-norm $|\psi|$ of $\psi\in L^2(\R^{3 N}; \, \HH')$ belongs to $L^{2}(\R^{3 N})$ and, by \eqref{QED semigroup inequality},
\begin{align}  
|e^{f}e^{-t H}e^{-f} \psi| \leq e^{f}e^{-t S}e^{-f}|\psi|.
\label{weighted pointwise ineq}
\end{align} 
We conclude that 
\begin{align}  
e^{f}e^{- t H}e^{-f}: L^2(\R^{3 N}; \, \HH') \longrightarrow L^{\infty}(\R^{3 N}; \, \HH')
\label{H mapping property}
\end{align} 
as a bounded linear operator. This allows for a short proof of the theorem below.
Notice that the class of potentials considered in \cite{Si82, M16, DHSV} is more general than \eqref{potential}. 
\bigskip

%----------------------------------------------------------------------------------------------------------------
\begin{Theorem}
\label{thm:ptw-decay-confined}
Let $H^{-}$ be the Pauli-Fierz Hamiltonian and $P_\lambda^{-} = \rchi_{(- \infty, \lambda]}(H^{-})$.
If $\lambda < \Sigma$ and $\beta < \sqrt{\Sigma - \lambda}$, then
\begin{align*}  
e^{\beta |\cdot|} P_\lambda^{-}: \HH^{-} \longrightarrow L^\infty(\R^{3N}; \, \HH')
\end{align*} 
is bounded. In particular, for all $\psi \in \mathrm{Ran} \, P_\lambda^{-}$ 
\begin{align*}  
\esssup_{x \in \R^{3 N}} e^{\beta |x|}|\psi(x)| < \infty. 
\end{align*} 
\end{Theorem}
%----------------------------------------------------------------------------------------------------------------

\noindent \emph{Remark:}
Both, the mapping property \eqref{H mapping property} and the result of this theorem are due to Matte \cite{M16}, see also Section 2.5 of this paper. He obtained these results by means of the Feynman-Kac representation of the semigroup of $H$, very much like the corresponding results for $S$ were established by Simon  \cite[Section B6]{Si82}. The point of our derivation is that, thanks to the semigroup version of  Kato's inequality, \Cref{Simon's theorem}, the machinery of stochastic integration can be avoided in favor of a much shorter and more direct argument.

%----------------------------------------------------------------------

\begin{proof}
Since $P_\lambda^{-}$ is a projector and commutes with $H^{-}$ we have
\begin{align*}  
e^{\beta|\cdot|} P_\lambda^{-} = \left[e^{\beta|\cdot|}e^{-t H}e^{-\beta|\cdot|}\right] \, \left[ e^{\beta|\cdot|} P_\lambda^{-} \right] \, \left[ e^{t H^{-}}P_\lambda^{-}\right].
\end{align*} 
Notice that $e^{t H^{-}}P_\lambda^{-}$ and $e^{\beta|\cdot|} P_\lambda^{-}$ are bounded operators on $\HH^{-}$. The latter follows from $\beta < \sqrt{\Sigma - \lambda}$ and \eqref{L 2 operator localization}. From \eqref{H mapping property} it follows that 
\begin{align*}  
e^{\beta|\cdot|} e^{-t H}e^{-\beta|\cdot|}: \HH = L^2(\R^{3 N}; \, \HH') \longrightarrow L^\infty(\R^{3 N}; \, \HH')
\end{align*} 
is bounded. This concludes the proof. 
\end{proof}

%--------------------------------------------------------------------------------------------------------------------------------------------------------------------------------------------------------------------------------------
\noindent
\textit{Acknowledgement.} V.K. thanks Heinz Siedentop and Simone Rademacher for the opportunity to present the first version of this paper at the DMV Meeting 2023 at TU Ilmenau. We thank Volker Bach, Dirk Hundertmark and Oliver Matte for pointing out relevant literature. Oliver Matte, in addition, sent us helpful comments on the second version. This work is supported by the Deutsche Forschungsgemeinschaft (DFG, German Research Foundation) - Projektnummer 531147062.

\appendix
\section{Appendix}
\label{sec:appendix}

\begin{Lemma}
$\mathrm{(Product \, rule)}$ Let $\Omega \subset \R^n$ be open. Suppose that $u, v \in L_\mathrm{loc}^1(\Omega)$ are weakly differentiable. If $u v \in L_\mathrm{loc}^1(\Omega)$ and $u (\nabla v) + (\nabla u) v \in L_\mathrm{loc}^1(\Omega)$, then $u v$ is weakly differentiable and
\begin{align*}
\nabla ( u v ) = u (\nabla v) + (\nabla u) v.
\end{align*} 
\label{Produktregel}
\end{Lemma}
\vspace{-7mm}
For the proof see \cite[Lemma 2.14]{FLW}. In our applications of this result, one of the two factors $u,v$ is bounded, which is the easier case established first in \cite{FLW}.

\begin{Lemma}
$\mathrm{(Chain \, rule)}$   
Let $f \in C^1(\R)$ and suppose that $f'$ is bounded. Let $\Omega \subset \R^n$ be open. If $u \in L_\mathrm{loc}^1(\Omega)$ is real-valued and weakly differentiable, then the composition $f \circ u$ is weakly differentiable and 
\begin{align*}
\nabla (f \circ u) = (f' \circ u) \, \nabla u. 
\end{align*} 
\label{Kettenregel}
\end{Lemma}
\vspace{-7mm}
For the proof see \cite[Theorem 7.8]{GT}.

\begin{Lemma}
\label{lower bound on spin}
For all $\psi \in D(H_f)$ we have that pointwise on $\R^{3 N}$
\begin{align*} 
(\psi, ( H_f +  \sigma \cdot B) \psi) \geq - \frac{8 \pi}{3} \alpha N^2 \Lambda^3 | \psi |^2.
\end{align*} 
\end{Lemma}

See, e.g.,  \cite[Lemma A.1]{LL} for the proof, but notice that the magnetic field operator in \cite{LL} differs by a factor of $(2 \pi)^{-1}$ from our definition.

%----------------------------------------------------------------------------------------------------------------------------------------------

\end{document}